\documentclass[preprint,5p]{elsarticle}
\usepackage{graphicx}
\usepackage{amssymb,amsmath,mathrsfs}
\journal{Physics Letter B}

\begin{document}
\begin{frontmatter}
\title{Shell Model description of the $\beta \beta$ decay of   $^{136}$Xe}

\author[STB]{E. Caurier}
\author[STB]{F. Nowacki}
\author[UAM]{A. Poves}

\address[STB]{IPHC, IN2P3-CNRS/Universit\'{e} Louis Pasteur
BP 28, F-67037 Strasbourg Cedex 2, France}
\address[UAM]{Departamento de F\'{\i}sica Te\'{o}rica,  Universidad Aut\'{o}noma
de Madrid and  Instituto de F\'{\i}sica Te\'{o}rica, UAM/CSIC, E-28049, Madrid, Spain}

\date{\today}
%\maketitle

\begin{abstract}   
We study in this letter the double beta decay of $^{136}$Xe with  emission of two
neutrinos which has been recently measured by the EXO-200 collaboration. We use the 
same shell model framework, valence space, and effective interaction that we have already  employed in
our calculation of the nuclear 
matrix element (NME)  of its neutrinoless double beta decay.   Using  the quenching factor of  the Gamow-Teller operator  which 
is needed to reproduce the very recent
high resolution $^{136}$Xe ($^{3}$He, $t$)$^{136}$Cs data, we obtain a nuclear matrix element
M$^{2\nu}$=0.025~MeV$^{-1}$ compared with the experimental value M$^{2\nu}$=0.019(2)~MeV$^{-1}$.

\end{abstract}

\begin{keyword} 
Shell Model, Double beta decay matrix elements
\PACS 21.10.--k, 27.40.+z, 21.60.Cs, 23.40.--s
\end{keyword}

\end{frontmatter}

 The  double beta decay is a rare process (second order in the weak interaction)
 which takes place when the single beta decay of the parent even-even nucleus to the neighbor
odd-odd nucleus is forbidden by energy conservation or highly suppressed by the angular momentum selection rules.  In addition it is one of the major
 sources of background for the even rarer  neutrinoless decay which, if detected,  will settle the nature (Majorana or Dirac)
 and the mass scale of the neutrinos. Until the EXO-200 measure \cite{exo:2011}  $^{136}$Xe  was the only
 (experimentally relevant) potential neutrinoless emitter whose two neutrino decay was unknown. In addition, the lower bound to its
 half life published by the Dama collaboration \cite{dama}  demanded  a nearly  complete cancellation of the
 nuclear matrix element. After the EXO measure, we know that the matrix element is small (indeed, the smallest
 among the measures ones)  but not pathologically so (see in Table \ref{tab1} the present status of the 2$\nu$
 decays from the recent compilation of ref. \cite{barab}). The EXO-200 measure has been confirmed by KamLAND-Zen
 \cite{kamland} only a few weeks ago.
 
The 2$\nu$ decay half-life
contains a phase space factor and the square of a nuclear matrix element

\begin{equation}
[T^{2\nu}_{1/2}]^{-1} = G_{2\nu} |M^{2\nu}_{GT}|^2 
\label{eq_t1/2}
\end{equation}
 
    The nuclear structure information is contained in the nuclear matrix element to which
only the Gamow-Teller $\sigma t^{\pm}$ part contributes in the long wavelength
approximation.

\begin{equation}
M^{2\nu} = \sum_m \frac { \langle 0^+_f | \vec{\sigma}  t^+ | m \rangle
\langle m | \vec{\sigma} t^+ | 0^+_i \rangle}{E_m - (M_i + M_f)/2} ~
\label{eq_m2nu}
\end{equation}
 
 Therefore, to calculate the nuclear matrix element we need to describe properly the ground state of the parent 
 and grand daughter nuclei as well as
 all the 1$^+$ excited states of the intermediate odd-odd nucleus.
In other words, the GT$^-$ strength function 
of the parent, the GT$^+$ strength function of the grand daughter  and  the relative
phases of the contributions from each intermediate state.

\begin{table}\begin{center}
     \caption{Experimental  2$\nu$  $\beta \beta$ decay matrix elements \label{tab1}}
   \begin{tabular*}{\linewidth}{@{\extracolsep{\fill}}lcc}
     \hline \\
Decay &  M$^{(2\nu)}$  (MeV$^{-1}$)  &  T$^{2\nu}_{1/2}$(y)  \\ [5pt] \hline \\  
 $^{48}$Ca $\rightarrow$  $^{48}$Ti  & 0.047$\pm$0.003 &   4.4 x 10$^{19}$ \\
 $^{76}$Ge $\rightarrow$  $^{76}$Se  & 0.140$\pm$0.005 &   1.5 x 10$^{21}$  \\
 $^{82}$Se $\rightarrow$  $^{82}$Kr  & 0.098$\pm$0.004 &    9.2 x 10$^{19}$\\
 $^{96}$Zr $\rightarrow$  $^{96}$Mo  & 0.096$\pm$0.004 &    2.3 x 10$^{19}$ \\
 $^{100}$Mo $\rightarrow$  $^{100}$Ru  & 0.246$\pm$0.007 &   7.1 x 10$^{18}$ \\
 $^{116}$Cd $\rightarrow$  $^{116}$Sn  & 0.136$\pm$0.005 &    2.8 x 10$^{19}$ \\
 $^{128}$Te $\rightarrow$ $^{128}$Xe & 0.049$\pm$0.006 &     1.9 x 10$^{24}$ \\
 $^{130}$Te $\rightarrow$ $^{130}$Xe & 0.034$\pm$0.003 &     6.8 x 10$^{20}$ \\
 $^{136}$Xe $\rightarrow$ $^{136}$Ba & 0.019$\pm$0.002 &    2.1 x 10$^{21}$\\
$^{150}$Nd $\rightarrow$ $^{150}$Sm &   0.063$\pm$0.003 &    8.2 x 10$^{18}$  \\
\hline\\
\end{tabular*}
\end{center}
\end{table}

 Our description of the wave functions of the states involved in the process is based in the Interacting Shell Model
 approach. The valence space includes the orbits  0g$_{7/2}$,
1d$_{5/2}$, 1d$_{3/2}$, 2s$_{1/2}$, and 0h$_{11/2}$, covering the sector of the nuclear chart  between N,Z=50  and N,Z=82.
We use the effective interaction gcn50:82 \cite{gcn:10} which is based in a renormalized G-matrix obtained 
from the Bonn-C \cite{bonnc} potential using the methods of ref. \cite{morten}. The final interaction is obtained
through a (mainly monopole) fit to about 300 energy levels from $\sim$90 nuclei in the region with  a rms deviation of 100~keV.
More details can be found in ref. \cite{xe110}.

 It is well known that the effective Gamow-Teller operator $\vec{\sigma} \;  t^{\pm}$ for  complete harmonic oscillator valence spaces
 can be approximated by  $ q \cdot \vec{\sigma} \; t^{\pm}$. $q$ is called the quenching factor and behaves as a sort of 
 effective GT charge (see ref.~\cite{javier} for a recent update of this topic). The value of $q$ has been fitted throughout the
  nuclear chart and the resulting values are 0.82, 0.77, and 0.74
 for the p, sd and pf shells.  Asymptotically  it tends to  0.7.  With the quoted value of $q$ for the pf-shell the half-life of the 2$\nu$  double
 beta decay of $^{48}$Ca \cite{ca48} could be predicted in perfect agreement with the later measured value \cite{ca48exp}.
 The problem arises when we try to describe heavier emitters  in which the minimal complete valence spaces (in the harmonic oscillator sense) are 
 still out of reach computationally.  A possible solution is to carry out a fit to all the experimentally available GT decays.
 We did this exercise in our valence space and the resulting value was $q$=0.57.
     A more accurate way of estimating the quenching factor is by comparing the theoretical predictions for the Gamow Teller strength
     functions relevant for the process with the experimental results obtained in charge exchange reactions. These data were not
     available for the   $^{136}$Xe  case until the advent of the results of  the  $^{136}$Xe  ($^{3}$He, $t$)$^{136}$Cs  reaction in the appropriate
     kinematics, which have been published very recently \cite{frekers}.  These results impact in our calculations in two ways; first
     because they give us the excitation energy of the first 1$^+$ state in $^{136}$Cs, 0.59~MeV, unknown till now, which appears in the energy denominator of
     equation \ref{eq_m2nu},  and secondly because it makes it possible to extract directly the quenching factor adequate for this process.

\begin{figure}[h] 
 \begin{center}
    \includegraphics[width=1.1\columnwidth]{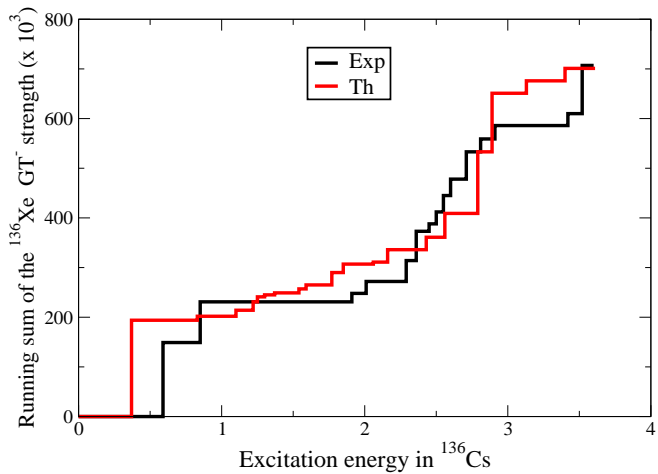}
\caption{\label{gt:rs}(color online) The running sum of the Gamow-Teller strength of $^{136}$Xe (energies in MeV). 
The theoretical strength is normalized to the experimental one.}
  \end{center}
\end{figure}

In what follows, we use for the A=136 isobars the wave functions that result
 of the large scale shell model calculations in the same valence space and with the same effective interaction which 
 we had  used in our calculation of the 0$\nu$ matrix elements of   $^{124}$Sn,  $^{128}$Te,  $^{130}$Te and $^{136}$Xe
 in ref. \cite{0nu}. First,
 we compare in Fig.~\ref{gt:rs}  the theoretical running sum of the B(GT$^-$) strength of  $^{136}$Xe
 with the experimental data from \cite{frekers}. We have normalized the total theoretical strength in the
 experimental energy window to the measured one. This implies a quenching q=0.45. Notice the very good agreement
 between the theoretical and experimental strength functions. If we had shifted the theoretical position of the first
 excited state of  $^{136}$Xe to its experimental value, the quenching factor would have been slightly larger.

\begin{figure}[h] 
 \begin{center}
    \includegraphics[width=1.1\columnwidth]{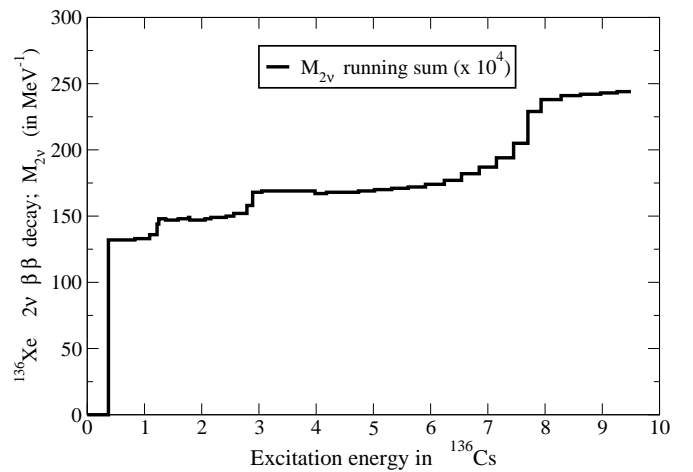}
\caption{\label{2nu:rs} The running sum of the  2$\nu$ matrix element of the double beta decay of $^{136}$Xe (energies in MeV).}
  \end{center}
\end{figure}

 Then we compute the 2$\nu$ matrix element  with the quenching factor extracted above. The result is given in Figure \ref{2nu:rs} in the form of a
 running sum. The final matrix element M$^{2\nu}$=0.025~MeV$^{-1}$ agrees nicely  with the experimental
 value. However,  one should bear in mind that the absolute normalization
 of the Gamow-Teller strength extracted from the charge exchange reactions may be affected by systematic errors, which
 could lead to modifications of the extracted quenching factor.  Minor variants of the gcn50:82 interaction which locally 
 improve the quadrupole properties of $^{136}$Ba
 lead to q=0.48 and M$^{2\nu}$=0.021~MeV$^{-1}$. Comparing Figure \ref{gt:rs} and Figure \ref{2nu:rs} it is evident
 that the  nuclear matrix element M$^{2\nu}$  does not saturate in the experimentally studied energy window. In fact,
 according to the calculation,  about 40\%  of  the total 2$\nu$ matrix element comes from states above it.
   
 For completeness, we present in Table~\ref{tab:ism} a compilation of the  2$\nu$ matrix element  elements for which 
 there are large scale shell model calculations. For the results of QRPA-like calculations see refs.~\cite{rpa1,rpa2,rpa3,rpa4}.
 \begin{table}\begin{center}
     \caption{The ISM predictions for the matrix element of  several 2$\nu$ double beta decays (in MeV$^{-1})$. 
     See text for the definitions of the  valence spaces and interactions. \label{tab:ism}}
   \begin{tabular*}{\linewidth}{@{\extracolsep{\fill}}ccccc}
     \hline \\  
       &   M$^{2\nu}$(exp) & q  & M$^{2\nu}$(th)    & INT \\ [5pt] \hline \\  
 $^{48}$Ca $\rightarrow$  $^{48}$Ti  & 0.047$\pm$0.003 & 0.74 & 0.047  & kb3  \\
 $^{48}$Ca $\rightarrow$  $^{48}$Ti  & 0.047$\pm$0.003 & 0.74  &  0.048 & kb3g  \\
 $^{48}$Ca $\rightarrow$  $^{48}$Ti  & 0.047$\pm$0.003 & 0.74  & 0.065 & gxpf1  \\
 $^{76}$Ge $\rightarrow$  $^{76}$Se  & 0.140$\pm$0.005 & 0.60  & 0.116 & gcn28:50   \\
 $^{76}$Ge $\rightarrow$  $^{76}$Se  & 0.140$\pm$0.005 & 0.60  & 0.120 & jun45  \\
 $^{82}$Se $\rightarrow$  $^{82}$Kr  & 0.098$\pm$0.004 & 0.60 & 0.126 & gcn28:50   \\
 $^{82}$Se $\rightarrow$  $^{82}$Kr  & 0.098$\pm$0.004 & 0.60  &  0.124 & jun45   \\
 $^{128}$Te $\rightarrow$ $^{128}$Xe & 0.049$\pm$0.006 & 0.57  & 0.059 & gcn50:82   \\
 $^{130}$Te $\rightarrow$ $^{130}$Xe & 0.034$\pm$0.003 & 0.57  & 0.043  & gcn50:82\\
 $^{136}$Xe $\rightarrow$ $^{136}$Ba & 0.019$\pm$0.002 & 0.45  & 0.025  & gcn50:82  \\
\hline\\
\end{tabular*}
\end{center}
\end{table}
In the calculation of the decay of  $^{48}$Ca, the valence space is the full $pf$ shell and the interactions are defined in refs. \cite{kb3} (kb3),
  \cite{kb3g} (kb3g),  and \cite{gxpf1} (gxpf1). The details can be found in refs.~\cite{ca48,horie}. The quenching 
  factor comes from a fit to all the experimentally available Gamow-Teller decays in the region using the kb3 interaction \cite{q_gab}.
  No equivalent fits are available for kb3g or gxpf1,  therefore we use the same quenching factor for all of them. We are
  convinced that the differences would be negligible in this case.  
  
 For the decays of  $^{76}$Ge and  $^{82}$Se we take a core of $^{56}$Ni and the valence space spanned by the orbits
1p$_{3/2}$, 0f$_{5/2}$, 1p$_{1/2}$, and 0g$_{9/2}$. The interactions are defined in refs. \cite{gcn28:50} (gcn28:50)  and
 \cite{jun45} (jun45). There is a published calculation with a preliminary version of jun45 in ref.~\cite{2nu_jun45}.  The value of
 the quenching factor q=0.60 was obtained in this reference from a fit to the Gamow-Teller decays in the region. As there
  are no fits available with the other interactions we adopt this value for them all.
The results for the 2$\nu$ NME compare quite well with the experimental values and are very much interaction independent, 
as in the $^{48}$Ca decay.

\begin{figure}[h] 
 \begin{center}
    \includegraphics[width=1.1\columnwidth]{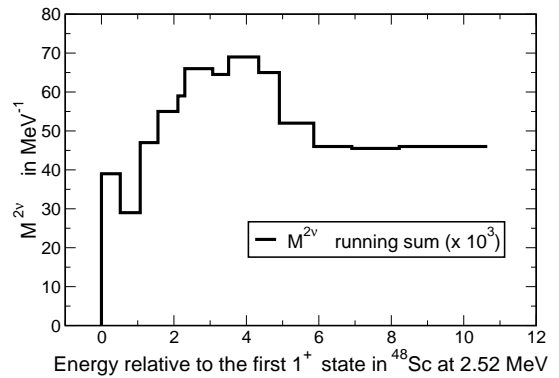}
\caption{\label{fig:sfa48} The running sum of the  2$\nu$ matrix element of the double beta decay of $^{48}$Ca (energies in MeV).}
  \end{center}
\end{figure}

\begin{figure}[h] 
 \begin{center}
    \includegraphics[width=1.1\columnwidth]{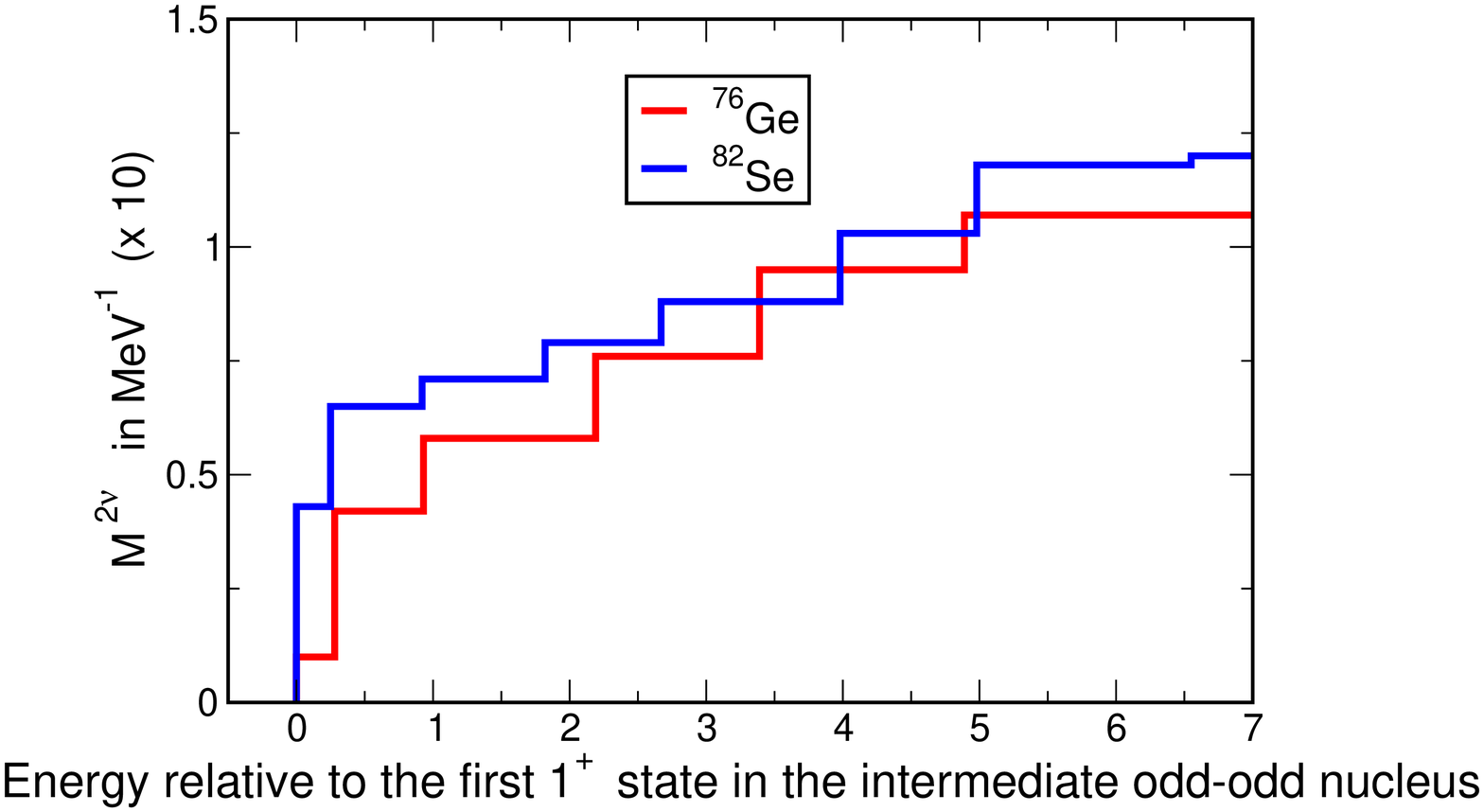}
\caption{\label{fig:sfa76}(color online) The running sum of the  2$\nu$ matrix element of the double beta decays of $^{76}$Ge (red) and $^{82}$Se (blue)(energies in MeV).}
  \end{center}
\end{figure}

 In addition,  we have plotted  in Figure~\ref{fig:sfa48}
 the running sum of  the matrix element of  the 2$\nu$ double beta decay  of $^{48}$Ca, using the kb3 interaction, and
   in Figure~\ref{fig:sfa76} the same for the decays  of $^{76}$Ge and $^{82}$Se
  using  the gcn28:50 interaction.  Notice that  the patterns in the three case are quite different and are also at variance with the results for the
  $^{136}$Xe decay shown in Figure  \ref{2nu:rs}, reflecting the different structures of the intermediate nuclei.
    The  $^{48}$Ca decay is peculiar in one aspect, to be the only case in which there are
  important canceling contributions to the  2$\nu$ NME. In fact,  if the  contributions from all the states would have had 
  the same sign,  the matrix element would have more than doubled, thus needing a quenching factor q=0.53 instead
  of q=0.74, in line with our findings in the other valence spaces. In the remaining cases the sign coherence is nearly complete. 
  However, it is difficult to make a more precise surmise 
  about the extra quenching factor needed when incomplete valence spaces (in the harmonic oscillator sense) are used,
  based solely in this example. Concerning the energy at which the 2$\nu$ NME  saturates,  the differences among the four
  cases studied are relatively minor, and one can safely conclude that  states beyond 10 MeV excitation  energy  above 
  the first 1$^+$ state in the intermediate 
  odd-odd nucleus do not contribute to  M$^{2\nu}$.
       
Finally, for the decays of  $^{128}$Te and  $^{130}$Te  we use the same valence space and interaction than
for the decay of $^{136}$Xe.  The quenching
 factor, as we have already mentioned, is obtained by a fit to the single Gamow-Teller decays experimentally known
 in this region (and within our computational limits).  Had we used the newer value from the analysis of the  
$^{136}$Xe charge exchange data, the values of the 2$\nu$ matrix elements would have been:

\medskip

M$^{2\nu}$($^{128}$Te)=0.037~MeV$^{-1}$  

M$^{2\nu}$($^{130}$Te)=0.027~MeV$^{-1}$

\medskip

In conclusion, we have shown that large scale shell model calculations which describe in detail the
spectroscopic properties of large regions of the nuclear chart can also make accurate predictions (or postdictions)
of the nuclear matrix elements of weak processes including the rarest ones; the 2$\nu$ and 0$\nu$ double beta decays.
In the  2$\nu$ case we have found that we can explain the experimental data provided the Gamow-Teller operator
is renormalized (quenched) so as to reproduce the Gamow Teller single beta decays or the charge exchange results
in the relevant regions. We have highlighted the case of $^{136}$Xe, whose  2$\nu$ double beta decay half-life has been
recently measured by the EXO-200 collaboration and confirmed by KamLAND-Zen.

{\bf Acknowledgements} Partially supported by the MICINN (Spain) 
                       (FIS2009-13377); by the IN2P3(France) and MICINN(Spain) (AIC10-D-402);
                       and by the Comunidad de Madrid (Spain) (HEPHACOS S2009-ESP-1473).


\begin{thebibliography}{99}

\bibitem{exo:2011}  N. Ackerman,  {\it et al.}  (EXO Collaboration)
                                 Phys. Rev. Lett. {\bf 107} (2011)  212501. 
                                 
\bibitem{dama}   R. Bernabei, {\it et al.}  Phys. Lett. B {\bf 546} (2002) 23.

         
\bibitem{barab} A. S. Barabash,   Phys. Rev. C {\bf 81} (2010) 035501.

\bibitem{kamland}  A. Gando,   {\it et al.} (The  KamLAND collaboration). arXiv:1201.4664v1 [hep-ex] January 2012.



\bibitem{gcn:10}  A. Gniady, E. Caurier, and F. Nowacki  (unpublished)

\bibitem{bonnc}  R. Machleidt, F. Sammaruca and Y. Song, Phys. Rev. C   {\bf 53}  (1996) R1483.


\bibitem{morten} M. Hjorth-Jensen, T. T. S. Kuo, and E. Osnes,   Phys. Rep. {\bf 261}  (1995) 126.

\bibitem{xe110}    E. Caurier, F. Nowacki,   A. Poves, and K. Sieja,  Phys. Rev. C   {\bf 82} (2010) 064304.

\bibitem{javier}   J. Men\'endez,  D. Gazit,  and  A. Schwenk, Phys. Rev. Lett. {\bf 107} (2011) 062501.
 
\bibitem{ca48} E. Caurier, A. Poves  and A. Zuker, Phys. Lett. B {\bf 252} (1990) 13.

    
\bibitem{ca48exp} A. Balysh,  {\it et al.}  Phys. Rev. Lett. {\bf 77} (1996)  5186. 
                                 
\bibitem{frekers}   P. Puppe, D. Frekers,  {\it et al.}  Phys. Rev. C {\bf 84} (2011) 051305R.
                              
\bibitem{0nu} J. Men\'endez, A. Poves, E. Caurier and  F. Nowacki,  Nucl. Phys.   {\bf A 818} (2009) 139.

\bibitem{rpa1} J. Suhonen and O. Civitarese, Physics Reports  {\bf 300} (1998) 123.

\bibitem{rpa2} F. Simkovic, A. Faessler, V. Rodin, and P. Vogel, Phys. Rev. C {\bf 77} (2008) 045503.

\bibitem{rpa3}  R. Alvarez-Rodr\'{\i}guez,  {\it et al.}  Phys. Rev. C {\bf 70} (2004)  064309.

\bibitem{rpa4}   S. Singh, R. Chandra, P.K. Rath, P.K. Raina, and J.G. Hirsch,  Eur. Phys. J. A {\bf 33} (2007)  375.


\bibitem{kb3}  A. Poves  and A. Zuker,   Physics Reports  {\bf 70}  (1981) 235.

\bibitem{kb3g}   A. Poves, J. S\'anchez Solano, E. Caurier and F. Nowacki,   Nucl. Phys. A {\bf 694}  (2001) 157.


\bibitem{gxpf1}  M. Honma, T. Otsuka, B. A. Brown and T. Misuzaki, Phys. Rev. C {\bf 65} (2002) 061301.


\bibitem{horie}  M. Horoi,  S. Stoica, and B. A. Brown, Phys. Rev. C {\bf 75} (2007)  034303.


\bibitem{q_gab}    G. Mart\'{\i}nez-Pinedo, A. Poves, E. Caurier y A. Zuker,    Phys. Rev. C {\bf 53}  (1996) R2602.


\bibitem{gcn28:50}      E. Caurier, J. Men\'endez, F. Nowacki, and A. Poves,  Phys. Rev. Lett.  {\bf 100}  (2008) 052503. 

\bibitem{jun45}   M. Honma, T. Otsuka, T. Misuzaki and M. Hjort-Jensen,  Phys. Rev. C {\bf 80} (2009) 064323.


\bibitem{2nu_jun45}  M. Honma, T. Otsuka, T. Misuzaki and M. Hjort-Jensen,  Journal of Physics; Conf. Ser. {\bf 49} (2006) 45.




\end{thebibliography}
\end{document}